# Anomalous thermal oxidation of gadolinium thin films deposited on silicon by high pressure sputtering

M.A. Pampillón, P.C. Feijoo, E. San Andrés ⇑, M.L. Lucía, A. del Prado, M. Toledano-Luque

Departamento de Física Aplicada III (Electricidad y Electrónica), Universidad Complutense de Madrid, Madrid 28040, Spain

**Abstract**

Thin gadolinium metallic layers were deposited by high-pressure sputtering in pure Ar atmosphere. Subsequently, in situ thermal oxidation was performed at temperatures ranging from 150 to 750 C. At an oxidation temperature of 500 C the films show a transition from monoclinic structure to a mixture of monoclinic and cubic. Regrowth of interfacial $SiO_x$ is observed as temperature is increased, up to 1.6 nm for 750 C. This temperature yields the lowest interface trap density, 4 $10^{10}$ eV$^1$ cm$^2$, but the effective permittivity of the resulting dielectric is only 7.4. The reason of this low value is found on the oxidation mechanism, which yields a surface with located bumps. These bumps increase the average thickness, thus reducing the capacitance and therefore the calculated permittivity.

## 1. Introduction

The continuous advance of the microelectronic industry will require the transition to third generation high permittivity dielectrics in the near future. This generation should present a higher permittivity value than the $HfO_2$ that is currently widely used [1], while keeping band offsets higher than 1 eV (to avoid Schottky emission) [2] and with good thermal stability on Si (to avoid silicon oxide regrowth during backend processing) [3].

An interesting candidate is $Gd_2O_3$, with a reasonably large-k value (around 18), large bandgap (6 eV) [4] and chemical stability in contact with Si [5]. Additionally, $Gd_2O_3$ is not only promising for high performance MOSFETs grown on Si, but it is also interesting as inter poly dielectric in flash memories [6] and as gate dielectric on III–V substrates [7].

A straightforward method to deposit Gd oxides is reactive sputtering of a Gd target with a mixed $Ar/O_2$ atmosphere [6]. The drawback of this approach is that the plasma highly reactive excited oxygen molecules can oxidize the substrate during the first stages of dielectric growth [8]. This may impose a limit on the device scaling. In order to explore processing alternatives to this conventional approach, in this work we have tested another $Gd_2O_3$ growth route aiming to the minimization of interface regrowth. This approach is inspired on previous works by Hayashi et al. that studied several Hf oxidation routes [9]. In our work we have first deposited a thin layer of metallic Gd in pure Ar atmosphere. Afterwards, without extracting the sample from the chamber, the film was oxidized in $O_2$ atmosphere at temperatures up to 750 C. The goal is to determine if there is an oxidation temperature high enough to fully oxidize the Gd layer but without affecting the underlying Si substrate (in other words, without the regrowth of interfacial $SiO_x$).

## 2. Experimental

Gadolinium metallic thin films were deposited by High Pressure Sputtering (in the following, HPS) from a metallic Gd target (purity 99.9%, diameter 5 cm). The target was sputtered by the application of rf power (between 10 and 30 W) in a high purity Ar atmosphere. The range of deposition pressures studied was 0.25–1.0 mbar, much higher than conventional sputtering systems. The reason of this high pressure is twofold: first, to thermalize the sputtered species within a very short distance reducing damage to the samples, and second, to confine the plasma within a small neighborhood of the target, that enables a very compact design of the system [10]. With this system we have already studied several high-k materials, like $TiO_2$ [10], $HfO_2$ [11] or $Sc_2O_3$ [12].

Concerning the processing procedure, before deposition the system was evacuated to a base pressure of 7.5 $\times$ 10$^7$ mbar using a turbomolecular pump. High purity Ar was introduced into the chamber through a mass flow controller. The optimal combination of power and pressure was selected by glow discharge optical spectroscopy (GDOS). After the Gd thin film was sputtered and without exposing the sample to the atmosphere, all pumping ports were closed, leaving the chamber sealed. Then, the chamber was filled with pure $O_2$ up to a pressure of 1 bar, and the substrate holder temperature was ramped at 50 C/min up to the oxidation temperature (we studied temperatures between 150 and 750 C).

This temperature was kept for 60 min (shorter times yielded incomplete oxidation). Then the substrates were cooled down to room temperature and subsequently removed from the chamber. The maximum temperature of 750 C was chosen with the aid of Technology Computer Aided Design, (TCAD), simulations which implement the Massoud oxidation model [13]. These simulations showed that annealing bare-Si up to 650 C in $O_2$ atmosphere did not cause any significant $SiO_2$ regrowth. Also, oxidation at 750 C only caused 1 nm increase in thickness. In any case, these simulations are only used as a first crude approach for selecting the temperature range, since it is not known beforehand whether covering Si with metallic Gd inhibits or catalyzes $SiO_2$ regrowth.

The substrates used were 2" n-Si wafers, grown by the floating zone method with (1 0 0) orientation. The substrates were cleaned by the conventional RCA procedure [14]. Just before the introduction to the chamber, in order to etch the native oxide, the wafers were immersed in diluted HF (50:1) for 30 s, and the HF was washed by dipping in deionized water. Once the substrates are loaded and the system has reached base pressure, the bare Si samples are heated at 500 C for 5 min in vacuum for native oxide desorption. For X-ray and FTIR characterization high resistivity (200–1000 X cm) both sides polished uncompensated substrates were used. On the other hand, low resistivity (1.5–5 X cm) single side polished substrates were used for electrical characterization. Transmission Electron Microscopy (TEM) samples were fabricated by slicing the MIS devices before the final forming gas anneal.

The emission lines of the HPS glow discharge were registered by a Jobin Yvon H-25 monochromator attached to a photon counting system, measuring the plasma emission at wavelengths between 280 and 520 nm with a resolution of 0.1 nm. The sheet resistance of the oxidized $GdO_x$ layers was measured by a conventional 4-point probe method. The FTIR spectra were measured between 400 and 4000 cm$^1$ with a Nicolet Magna-IR 750 series II spectrometer. Glancing incident X-ray diffraction (XRD) was measured with a XPERT MRD Panalytical system for 2h angles between 10– 70. To calculate the thicknesses of the layers, the same system in a h-2h configuration was used to measure the X-ray reflection at low angles (XRR), between 0.2 and 4. The thicknesses were obtained by fitting the reflection spectra. Cross-sectional TEM images were obtained with a JEOL-JEM-2000FX microscope operating at 200 keV.

To fabricate MIS devices, 70 nm of Al was e-beam evaporated on top of the dielectrics and on the back-side of the wafer. Squares were defined by positive lithography and wet etch of Al. After that, the devices were annealed in forming gas (FGA) for 20 min at two typical processing temperatures, 300 and 450 C. The admittance of the resulting MIS devices was measured with an Agilent 4294A impedance meter. To correct for possible lithography offsets, several areas were measured (500  500 lm$^2$, 300  300 lm$^2$, and 100  100 lm$^2$). The interface trap density ($D_{it}$) was estimated by the conductance method. $C_{gate}$ vs $V_{gate}$ curves were fitted using the Hauser algorithm [15]. From the simulated Equivalent Oxide Thickness (EOT) and the measured physical thicknesses, the permittivity of the $GdO_x$ layer was calculated. Also, the leakage of the devices was measured with a Keithley 4200A system.

**3. Results and discussion**

The first objective was selecting the optimal sputtering conditions. We measured the GDOS spectrum of the plasma for rf power between 10 W and 30 W while setting the pressure at 0.5 mbar. The results are shown in Fig. 1a. As expected, it is observed that increasing power results in enhanced plasma activity. Also, no features related to $H_2O$ (typically a broad band between 310 and 320 nm [16]), nor $N_2$ peaks (the main peaks located at 336 and 358 nm [16]) were observed. This indicates that the chamber is correctly sealed, which is critical in order to ensure that no oxidation occurs during the Gd pulverization. The 10 W spectrum shows only ArI peaks (neutral Ar) between 415 and 430 nm [17] with a low intensity. This indicates that at this low power there are no ionized species, which are needed for the sputtering to take place. Increasing the power to 20 W increases the ArI intensity, and also many peaks make apparition: ArII (ionized Ar) is found in the 430– 490 nm range, some GdI peaks (the most intense one located at 422.5 nm) and also many GdII emission lines (in the 280–400 nm range) [18]. The effect of further increasing the power to 30 W is an increase on peak intensities, more relevant for the GdII case, but no new peaks appear. Thus we have selected 30 W as working power, in order to avoid being close to the extraction threshold power (which lies between 10 and 20 W, as we have found) and also to have a high deposition rate.

In order to determine the optimal working pressure, the plasma spectra were measured at pressures of 0.25, 0.50, 0.75, and 1.0 mbar, and they are shown in Fig. 1b. Higher pressures produced plasma instabilities. From the figure it is clear that at a pressure of 0.25 mbar the plasma does not present ionized Ar, thus there is no Gd extraction. For higher pressures there is Gd extraction, and the Gd vs Ar intensity ratio is higher on 0.50 mbar, which indicates a more efficient extraction. Thus we chose this pressure, 0.50 mbar, as working pressure.

Once the deposition conditions were fixed, as a fast screening procedure we studied the oxidation degree of the metallic films by 4-point sheet resistance measurements, in order to obtain the onset temperature of oxidation. We performed consecutive annealing treatments at increasing temperature, and we found that under 350 C the sheet resistance was finite with an ohmic behavior. On the other hand, for higher temperatures the resistance was too large to be measured. This was a first indication that full oxidation was not achieved for temperatures under 350 C.

Thick Gd films (around 200 nm) were deposited and oxidized at temperatures between 150 and 750 C in order to study the crystalline structure of the oxidized Gd films. Thick layers were used in order to obtain intense diffraction peaks. The XRD diffraction patterns of as-deposited metallic Gd and films oxidized at different temperatures are shown in Fig. 2.

Metallic Gd films (Fig. 2a) have the expected hexagonal structure, with the main diffraction peaks for planes (1 0 0) at 28.3 and (1 0 1) at 32.3 [19]]. For oxidation temperatures between 150–250 C there are traces of hexagonal Gd peaks (Fig. 2b). These residual metallic peaks disappear at higher temperatures. Also, after oxidation at low temperature many peaks make apparition. The most intense peak is located at 29.5, and its height is directly correlated to another peak at 26 As it is shown in Fig. 2, most peaks can be directly related to the diffraction of the monoclinic structure of $Gd_2O_3$ [20]. After oxidation at 350 C metallic Gd peaks disappear and the only phase present is monoclinic gadolinium oxide, as it is shown in Fig. 2c. At 500 C and above two more peaks appear (Fig. 2d and e). These are located at 20 and 28.5, but their intensity is 2 to 3 times lower. These peaks may correspond to the cubic $Gd_2O_3$ structure [21].

Peaks at 26, 29.5, and 33.9 could also be related to hexagonal $GdSi_2$ [22]. Several groups have reported the formation of $GdSi_2$ when Gd is deposited on Si [23]. The apparition of this silicide would happen before oxidation. However, silicides present ametallic character. Thus, if it were $GdSi_2$, the electric characteristic would be anomalous. However, this behavior is not observed, as we will see later in the C–V curves. Some peaks could correspond also to orthorhombic $Gd_2Si_2O_7$ [24], so gadolinium silicate could appear during oxidation. This fact would decrease the effective k value of the stack since it has a very low k value.

These results confirm the resistivity measurements that pointed to a minimum oxidation temperature of 350 C. Also it can be concluded that oxidation at low temperature produces a polycrystalline monoclinic $Gd_2O_3$ and gadolinium silicate may be forming in the interface. On the other hand, oxidation at temperatures above 500 C yields a mixture with cubic phase. This behavior is contrary to thermodynamic studies and the results of other groups [20,25], which found the opposite trend: that the cubic phase was favored at low temperature, while annealing at higher temperatures promoted the transition to the monoclinic phase. It is noteworthy that the reported k value of the monoclinic phase is significantly lower than the cubic phase, so if the device processing requires high temperatures this anomalous behavior might be useful.

In order to check if these conclusions can be extended to thin films, 15 nm - thick films were also studied by XRD. This thickness was also the chosen one to fabricate the MIS devices. As expected, the measurements were much noisier and the only observed diffraction was monoclinic (401) $Gd_2O_3$ peak, one of the most intense, at 29.4 [26]. However, above 500 C the peak broadened towards lower angles, pointing to the same phase mixture as thick layers. Also, since the surface of these films is closer to the Si substrate, the XRD would show silicate diffraction peaks if there were gadolinium silicate formation close to the interface.

In order to obtain the thickness of the $Gd_2O_3$ layers, XRR measurements were also performed. The results were fitted with a segmented algorithm to a $Gd_2O_3/SiO_2/Si$ bilayer model, using density, thickness and roughness as parameters. As an example, Fig. 3 shows the measurements for the film oxidized at 750 C, and the fitting results. There is an excellent agreement between measurements and simulation, as it can be seen in the figure.

The thickness values obtained from the fitting are shown in Fig. 4. The uncertainty of these values can be estimated as ± 1 nm. This absolute uncertainty is relatively small compared to the $Gd_2O_3$ thickness. We can observe that there is a thickness increase when annealing above 350 C (the temperature where the monoclinic $Gd_2O_3$ phase appears). For temperatures above 500 C (cubic + monoclinic phase) there are no relevant changes on $Gd_2O_3$ thickness, with a mean value of 13.6 nm. On the other hand, the relatively uncertainty on $SiO_2$ thickness is too high to obtain accurate quantitative information. The only trend that is clear is that above 350 C there is an increase of the interfacial $SiO_2$ thickness.

Since one of the goals of this work is trying to minimize $SiO_2$ or gadolinium silicate regrowth, we relied on transmission FTIR spectroscopy as a fast and more accurate technique to study the $SiO_2$ presence on the samples. The drawback of this technique is that it measures the film and the substrate, thus a reference substrate is needed to correct the spectrum. Our experimental procedure was to measure a Si wafer of the same lot HF etched just before FTIR measurements, in order to minimize native oxide regrowth. This spectrum was used to correct for the sample substrate absorbance. However, this approach has two drawbacks: slight differences in wafer thicknesses yield residual Si peaks in the spectra and some native oxide growth is unavoidable because after HF etching the substrate is transferred to the FTIR system in an uncontrolled atmosphere. This has the effect that for these small thicknesses FTIR can only be used qualitatively for $SiO_x$ thickness determination.

Concerning $Gd_2O_3$, due to the high atomic mass of Gd, the vibration of the Gd–O bond would shift to wavenumbers around 405–455 $cm^1$ [27]. On the other hand, the Gd-O-Si bond would show an absorbance at 900 $cm^1$ [28].

Fig. 5 shows the absorbance in the 850–1250 $cm^1$ range. No other relevant features are found in the rest of the spectrum. The peaks at 883 and 960 $cm^1$ are due to Si vibration, in other words, to the substrate correction. In particular we have not found any clear vibration on 900 $cm^1$, so it is not likely that there is a silicate formation. In fact, most peaks in the spectra are identical in shape and intensity, and can be related to the Si substrate correction. The exception is the feature at around 1060 $cm^1$, which shows a clear increase in absorption with increasing oxidation temperature. This feature is due to the $SiO_x$ that is usually present at the High-k/Si interface. Unstressed thermal $SiO_2$ has its most intense stretching absorption at 1076 $cm^1$ [29]. On the other hand, in silicon-rich $SiO_x$ this absorption shifts towards lower wavenumbers with increasing Si content [10]. In our oxidized Gd samples we observe that at 350 C there is a peak downwards, at 500 C there is no peak and at higher temperatures the peak grows in intensity (at 750 C the peak is located at 1065 $cm^1$ and its intensity is comparable to the intensity presented by a RCA cleaned substrate, which is covered by a 2–3 nm thick $SiO_2$ layer). Also, with increasing temperature the Si-O stretching peak shifts towards the thermal $SiO_2$ value (1076 $cm^1$). The interpretation is quite straightforward: at 350 C there is even less interfacial $SiO_x$ than native $SiO_x$ in the surface of the HF cleaned substrate (during the reference substrate measurement some native oxide regrowth is unavoidable). For higher temperatures the thickness of the interfacial layer grows, but it is difficult to quantify numerically this regrowth. The shift of the peak maximum indicates a relaxation of the $SiO_2$ film.

The electrical results of the MIS devices are shown on Fig. 6, where we present representative ($C_{gate}$, $J_{gate}$) vs $V_{gate}$ characteristics. The devices oxidized at 350 C presented very poor electrical characteristics due to gate leakage, and are not shown. The measuring frequency was chosen by obtaining $C_{gate}$ vs frequency in accumulation in the 1 kHz–1 MHz range. The signal frequency was chosen as high as possible in order to maximize capacitance signal, but not too high to avoid capacitance decrease (due to the coupled effect of conductance and series resistance) [30]. Thus, the devices were measured at 10 and 100 kHz. No relevant differences were found, thus here we only show the 10 kHz curves. The spread in accumulation capacitance within each sample was about 5–10%.

On Fig. 6a first we can see that the maximum capacitance increases 50% when the oxidation temperature increases from 500 to 650 C, but higher temperatures do not further increase capacitance (in fact, there is a slight capacitance decrease). This is an indication that further increasing oxidation temperature is not a good path to improve device characteristics. Also for FGA at 450 C the $C_{gate}$–$V_{gate}$ traces are free of humps in the depletion region, which is indicative of a low density of interface states. This is not the case when the FGA is performed at 300 C. For that temperature the curves are displaced towards positive voltages, and also show a clear hump in the depletion region, due to interface defects. This last fact indicates that FGA temperature above 300 C is mandatory.

The EOT of the devices was obtained from the fit of the $C_{gate}$–$V_{gate}$ FG annealed at 450 C with the CVC algorithm [13], and the results are summarized on Table 1. The fit of the samples FGA at 300 C is less reliable due to the big $D_{it}$ humps, but the resulting values are identical within ±0.5 nm, which indicates that FGA temperature does not affect $SiO_2$ regrowth or $Gd_2O_3$ permittivity. From these EOTs using the XRR thicknesses for $Gd_2O_3$ and $SiO_2$ at each temperature, the permittivity of the $Gd_2O_3$ film can be calculated with a two capacitors in series model. As it is also shown in Table 1, the maximum $k_{GdO}$ value of 7.4 is obtained at oxidation temperatures above 650 C. Unfortunately, this value is way too low for high performance high-k applications.

Concerning the gate leakage, it is clear that oxidation at higher temperatures reduces leakage by several orders of magnitude. Also, FGA at 450 C also reduces leakage by an order of magnitude as compared to FGA at 300 C. This reduction can be attributed to an improvement of the quality of the $Gd_2O_3$ layer, probably due to grain size growth, and to the densification/relaxation/defect passivation of the $SiO_x$ interfacial layer (as we previously showed, with increasing temperature the Si-O stretching peak shifts towards the relaxed $SiO_2$ value).

Another improvement that happens with oxidation temperature is also evident when extracting the interface trap density ($D_{it}$) from the electrical measurements by the conductance method. The results for 10 kHz are also included on Table 1, both for 300 C and 450 C forming gas anneals. There it can be seen that FGA at 300 C is ineffective in passivating interface defects, with $D_{it}$ values above $10^{12}$ eV$^1$ cm$^2$. On the other hand, by annealing at 450 C the interface presents a very low trap density, comparable to thermal $SiO_2$ (4 $10^{10}$ eV$^1$ cm$^2$).

The main drawback that we have found is the moderate permittivity value found. The origin of this low permittivity could be extrinsic (a measuring problem, such a miscalculation of the thicknesses obtained from XRR) or intrinsic (silicate formation, low permittivity of the dominant monoclinic phase...). In order to clarify this question, we prepared cross-sectional TEM samples of the samples oxidized at 500 and 750 C. The results for the 750 C sample are shown in Fig. 7.

On Fig. 7a we can see that the thicknesses calculated from XRR are very close to the actual thicknesses, since for oxidation at 750 C the $Gd_2O_3$ film is 15.2 ± 0.2 nm thick, and the $SiO_2$ interface is 1.6 ± 0.1 nm (at 500 C the thicknesses are 15.0 ± 0.2 and 1.4 ± 0.1 nm respectively). If the $Gd_2O_3$ permittivity is calculated using these thicknesses, very similar values are found (Table 1).

However, there are other regions on the surface where the film presents "bumps" (Fig. 7b shows the same device with less magnification). These bumps were found for both temperatures. Also, the TEM image of this bumps suggest a dome (or igloo) structure. In other words, there seems to be an empty/less dense space between the $Gd_2O_3$ and the Si. The origin of these bumps is unclear: they might be due to Si–Gd intermixing (in other words, the formation of gadolinium silicate as we pointed out when discussing the XRD results), but the flatness of the Si surface and the previous FTIR results rule out this possibility. Also, a reaction of Al with $Gd_2O_3$ is not likely, since the TEM samples were prepared after e-beam Al evaporation but before FGA, so thermal reaction is not likely. Furthermore, no significant differences in EOT were found for the two FGA temperatures studied, so a thermally activated reaction can be discarded. On reference [31] Mólnar et al. found that when annealing Gd on Si, even at low temperatures (320 C), due to the free Gibbs energy of the system, an "explosive" reaction between Si and Gd happened, starting on the weak $SiO_2$ spots. At these spots, gadolinium silicide flakes made apparition with a fractal structure. On our samples a similar effect might be the origin of the bumps: during oxidation, first we fill the chamber with $O_2$ and then we increase temperature. The oxygen comes to the film from the top Gd surface, while heat flows from the Si substrate, so in the first stages of the oxidation some explosive Gd–Si flakes could be produced at the Si interface, that afterwards become oxidized and give rise to the bumps. Another possibility is that it is a stress problem during warm up/cool down, but since the amount and size of the domes at 500 and 750 C are similar, we think that this explanation is less likely.

In any case, independently on the origin of the bumps, they are clearly at the origin of the low effective permittivity calculated: even if the permittivity were the same as $Gd_2O_3$, since they are 2–3 times thicker and occupy a big portion of the MIS surface, they produce a big decrease on capacitance. This decrease would be even more exacerbated if the dome explanation is correct, since in that case the capacity of the bump is the $Gd_2O_3$ capacity in series with the vacuum capacity, which has a relative permittivity of 1.

As a consequence of this problem, even after the low $D_{it}$ values found, we have to conclude that pure thermal oxidation of metallic Gd does not meet the requirements for future high-k dielectrics. Thus, following a similar approach as the works on $HfO_2$ of Y. Hoshino et al. [32], we are currently studying plasma oxidation of metallic Gd layers, and we expect to overcome these problems soon.

## 4. Conclusion

In this paper we have explored the thermal oxidation of metallic Gd in order to obtain high-k $Gd_2O_3$ on Si. The results show that the films oxidized above 500 C are a mixture of monoclinic and cubic phases. The leakage and interface trap density decrease with temperature, being minimal after oxidation at 750 C and FGA at 450 C. However, we found a low effective permittivity of the MIS device, due to the formation of domes at the $Gd_2O_3$/Si interface. These bumps decrease capacitance by a large amount, so the advantage of the high-k $Gd_2O_3$ is lost. Thus, in the future we will explore other processing approaches in order to obtain device quality high-k $Gd_2O_3$.

**Acknowledgements**

The authors want to acknowledge CAI de Técnicas Físicas, CAI de Espectroscopía, CAI de Difracción de Rayos X, CAI de Microscopía, Plataforma de Nanotecnologia of the Parc Científic de Barcelona for sample preparation and characterization. This work was funded by the projects TEC2007-63318, TEC201018051, GR58/08 and by the MEC FPU program (AP2007-01157).preparation and characterization. This work was funded by the projects TEC2007-63318, TEC201018051, GR58/08 and by the MEC FPU program (AP2007-01157).

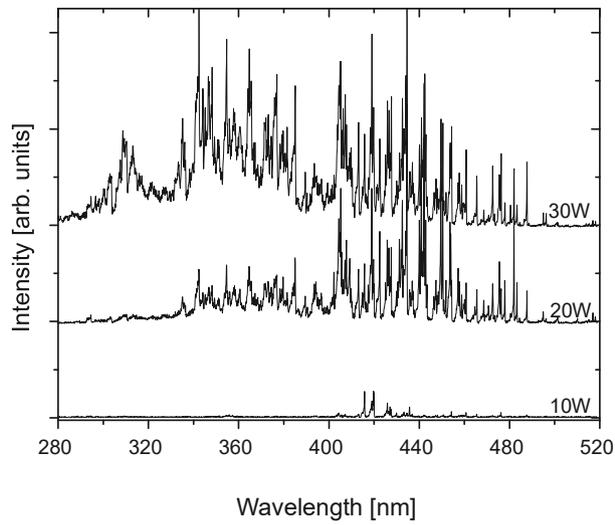

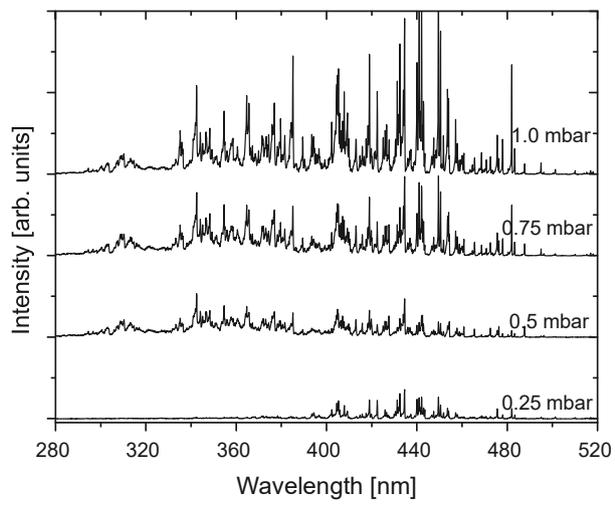

Fig. 1. Optical Spectra of Gd sputtering. Top: rf power variation (10–30 W). Down: Ar pressure dependence (0.25–1.0 mbar).

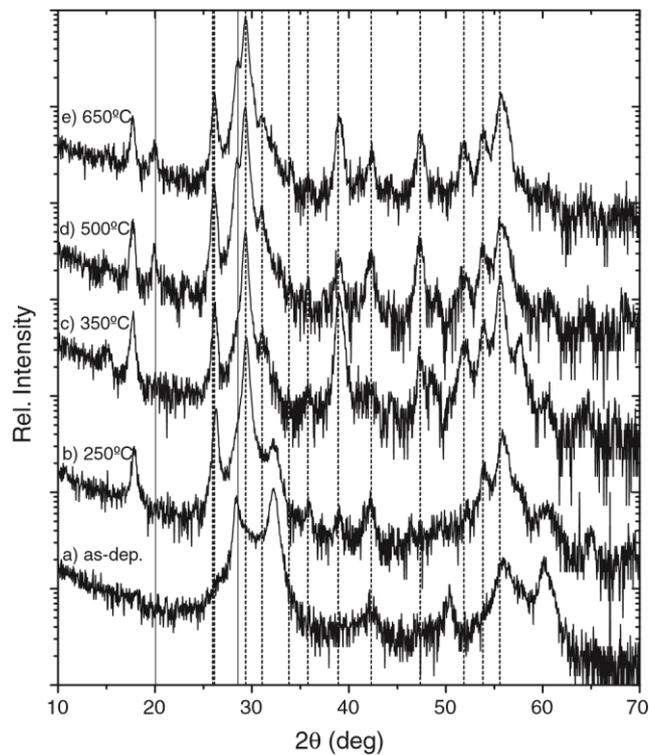

Fig. 2. X-ray diffraction patterns of thick Gd layers as-deposited (a) and oxidized at several temperatures (b–e). The observed reflection peaks have been identified as monoclinic gadolinium oxide (dashed lines) or cubic gadolinium oxide (solid lines).

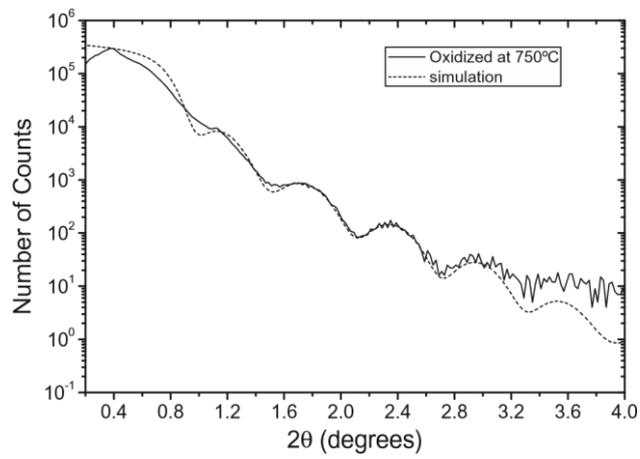

Fig. 3. X-ray reflection measurements of the Gd film oxidized at 750 C (solid line), and fit to a $Gd_2O_3/SiO_2/Si$ model (dashed line).

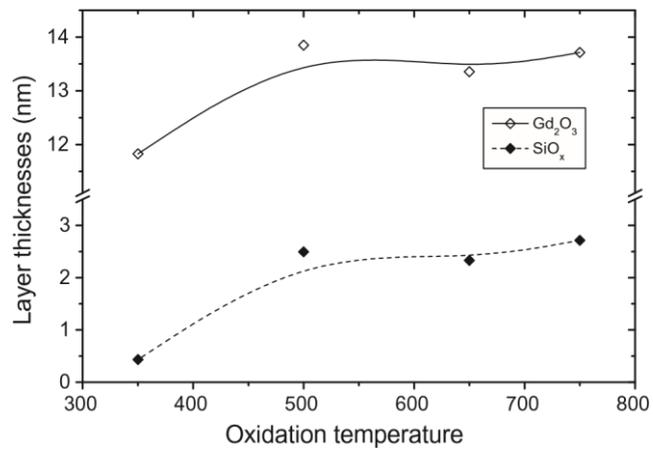

Fig. 4. Thickness result of the XRR simulations to a Gd$_2$O$_3$/SiO$_2$/Si bilayer model: Gd$_2$O$_3$ thickness (open symbols) and SiO$_2$ thickness (closed symbols).

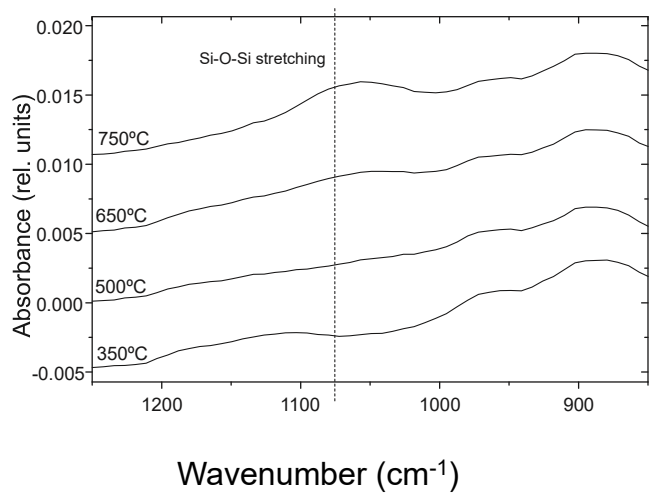

Fig. 5. FTIR spectra of thin Gd layers oxidized at several temperatures. The wavenumber of thermal SiO$_2$ stretching is marked by the dashed line.

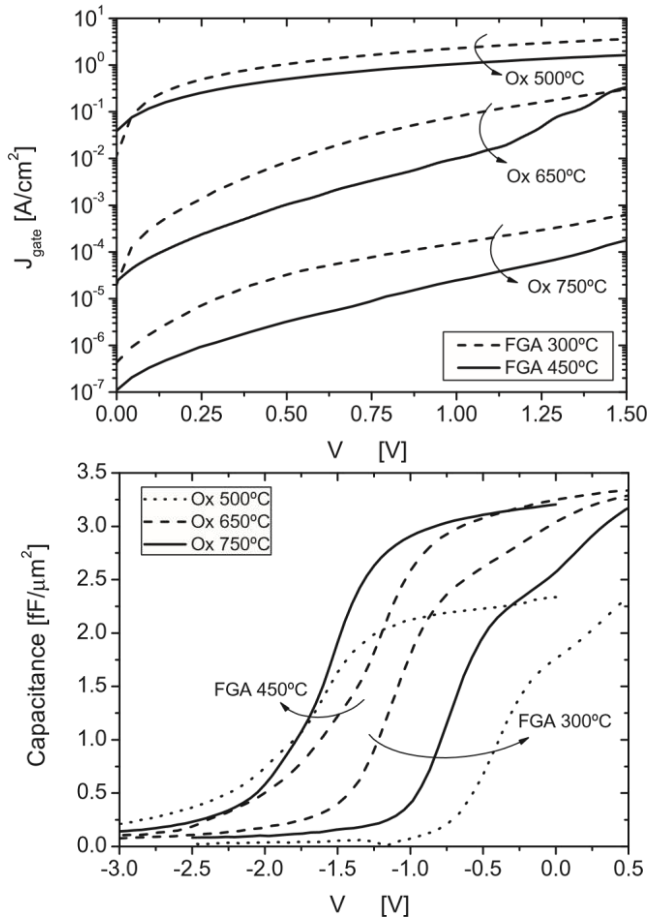

Fig. 6. Top) Representative gate capacitance curves as function of gate voltage of the Al/Gd$_2$O$_3$/SiO$_2$/Si devices measured at 10 kHz for different oxidation temperatures (500–750 C) and FGA temperatures (300–450 C). Bottom) Gate current density as a function of gate voltage of the same samples.

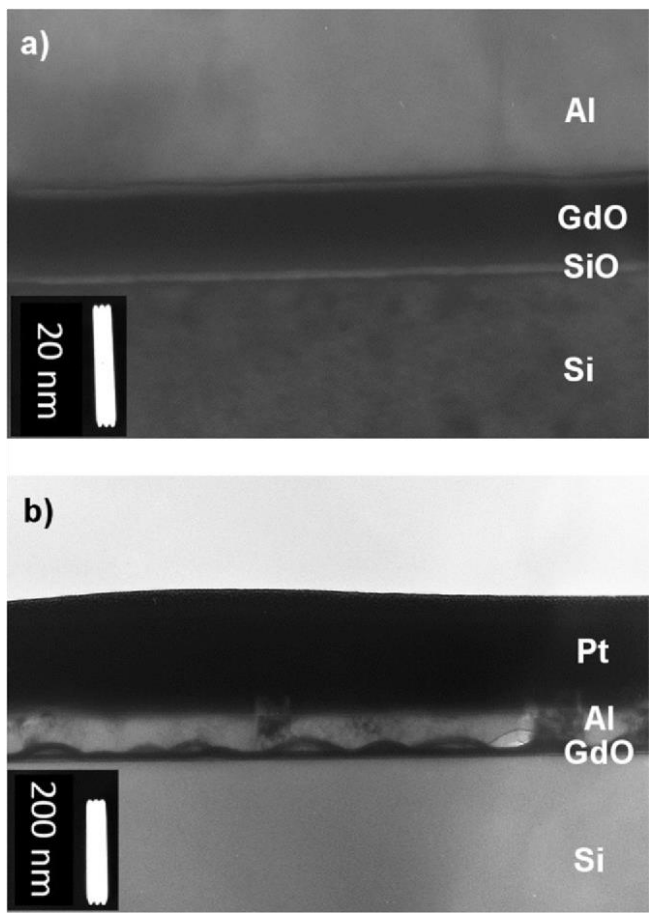

Fig. 7. Cross-sectional TEM images of the $Gd_2O_3$ films oxidized at 750 C: (a) high magnification image, used to measure $SiO_2$ thickness (b) low magnification image, where the density of domes can be observed. Samples were covered by Pt for protection during FIB sample preparation.

| Oxidation temperature (C) | EOT (nm) | $k_{GdO}$ (XRR thicknesses) | $k_{GdO}$ (TEM thicknesses) | $D_{it}$ (eV$^{-1}$ cm$^{-2}$) FGA 300 C | $D_{it}$ (eV$^{-1}$ cm$^{-2}$) FGA 450 C |
|---|---|---|---|---|---|
| 500 | 14.4 | 4.5 | 4.8 | >10$^{13}$ | >10$^{13}$ |
| 650 | 9.8 | 7.0 | 7.4 | 2 10$^{12}$ | 3 10$^{11}$ |
| 750 | 10.0 | 7.4 | 7.3 | 1.7 10$^{12}$ | 4 10$^{10}$ |

Table 1
Extracted electrical Parameters.